\documentclass[journal]{IEEEtran}

\usepackage{kotex}
\usepackage{graphicx}
\usepackage{amsmath}
\usepackage{algorithm}
\usepackage{algpseudocode}
\usepackage{tabularx}
\usepackage[table]{xcolor}
\usepackage{hhline} 
\usepackage{subcaption}
\usepackage{amsfonts}
\usepackage{cite}
\usepackage{color}
\usepackage{url} 
\usepackage{amssymb} 
\usepackage{booktabs}
\usepackage{array}      
\usepackage{multirow}   
\usepackage{tabularx}

\algnewcommand{\IIf}[1]{\State\algorithmicif\ #1\ \algorithmicthen}
\algnewcommand{\ELIIf}[1]{\State\algorithmicelse\ \algorithmicif\ #1\ \algorithmicthen}
\algnewcommand{\ElseIIf}[1]{\algorithmicelse\ #1} 
\algnewcommand{\EndIIf}{\unskip\ \algorithmicend\ \algorithmicif}

\begin{document}
\title{User-Centric Stream Sensing for Grant-Free Access: Deep Learning with Covariance Differencing}

\author{Sojeong Park, Yeongjun Kim, and Hyun Jong Yang,~\IEEEmembership{Senior Member,~IEEE}

\thanks{Sojeong Park is with the Department of Electrical Engineering, Pohang University of Science and Technology (POSTECH), Republic of Korea (e-mail: sojeong@postech.ac.kr).
Yeongjun Kim is with the System LSI Modem Development Team, Device Solutions, Samsung Electronics, Republic of Korea (e-mail: yj0531.kim@samsung.com).
Hyun Jong Yang is with the Department of Electrical and Computer Engineering and Institute of New Media and Communications, Seoul National University, Republic of Korea (email: hjyang@snu.ac.kr).
}
}

\maketitle

\begin{abstract}
Grant-free (GF) access is essential for massive connectivity but faces collision risks due to uncoordinated transmissions. While user-side sensing can mitigate these collisions by enabling autonomous transmission decisions, conventional methods become ineffective in overloaded scenarios where active streams exceed receive antennas. To address this problem, we propose a differential stream sensing framework that reframes the problem from estimating the total stream count to isolating newly activated streams via covariance differencing. We analyze the covariance deviation induced by channel variations to establish a theoretical bound based on channel correlation for determining the sensing window size. To mitigate residual interference from finite sampling, a deep learning (DL) classifier is integrated. Simulations across both independent and identically distributed flat Rayleigh fading and standardized channel environments demonstrate that the proposed method consistently outperforms non-DL baselines and remains robust in overloaded scenarios.
\end{abstract}

\begin{IEEEkeywords}
Grant-free access, massive MIMO, stream sensing, deep learning.
\end{IEEEkeywords}
\vspace{-3mm}

\section{Introduction}
The evolution toward sixth-generation networks is characterized by the convergence of massive machine-type communication and ultra-reliable low-latency communication \cite{6gsurvey1, 6gsurvey2, 6gsurvey3}. 
While massive multiple-input multiple-output (MIMO) systems facilitate massive connectivity by providing sufficient spatial degrees of freedom \cite{massivemimo1}, grant-based (GB) access remains a critical bottleneck. This is due to the multi-step exchange of scheduling requests and buffer status reports, which incurs significant signaling overhead and latency \cite{3GPP.38.321}.

Grant-free (GF) access has emerged as a compelling solution, allowing devices to transmit immediately without prior scheduling \cite{gf1, gf2, gf3}. However, its uncoordinated nature increases the risk of collisions and interference. To address this, user equipment (UE)-side sensing can support autonomous transmission decisions through listen-before-talk (LBT) strategy, where a UE senses the channel prior to transmission \cite{lbt}. This mechanism allows UEs to assess channel availability, thereby reducing network congestion.

While UE-side sensing provides a promising means to reduce collisions, its practical realization is constrained by hardware limitations. Unlike a base station (BS), a UE is typically equipped with a limited number of antennas. In dense environments, the number of active streams often exceeds the antenna count, resulting in an under-determined system. In such overloaded scenarios, conventional source enumeration methods such as minimum description length (MDL) fail because the spectral gap between stream and noise singular values becomes indistinguishable \cite{mdl}. Therefore, a more robust sensing methodology is required to maintain accuracy in mathematically ill-posed scenarios.

To overcome these challenges, we propose a user-centric stream sensing framework tailored to overloaded scenarios. Diverging from network-centric approaches, our method empowers individual UEs to make autonomous transmission decisions by locally sensing the uplink environment, thereby mitigating collision risks. Instead of estimating the total stream count, we reframe the objective to detect only newly activated streams since the previous observation. To ensure robust performance of this incremental tracking even in mathematically ill-posed regimes, we integrate deep learning (DL) into the framework. DL has emerged as a transformative paradigm for the wireless physical layer \cite{dl1, dl2, dl3}, capable of learning complex non-linear mappings directly from data \cite{dlbook}. In our context, DL enables the system to extract informative features in overloaded scenarios.
\begin{figure} [t]
    \centering
    \includegraphics[width=0.9\linewidth]{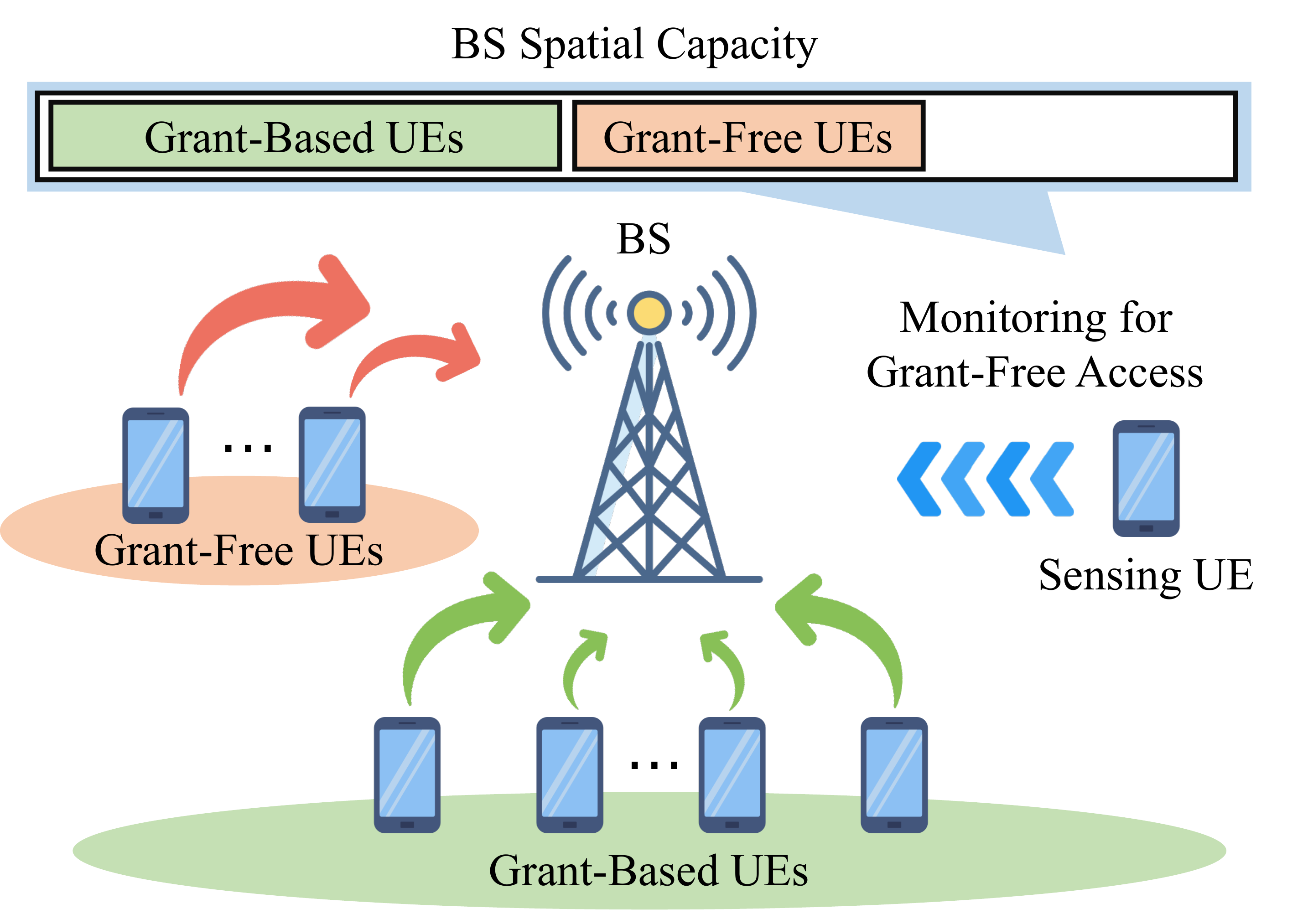}
    \caption{System model for user-centric stream sensing in a GF network by monitoring the number of active GF UEs.}
    \vspace{-5mm}
    \label{fig:system_model}
\end{figure}
\begin{figure*}[t!]
    \centering
    \includegraphics[width=0.95\linewidth]{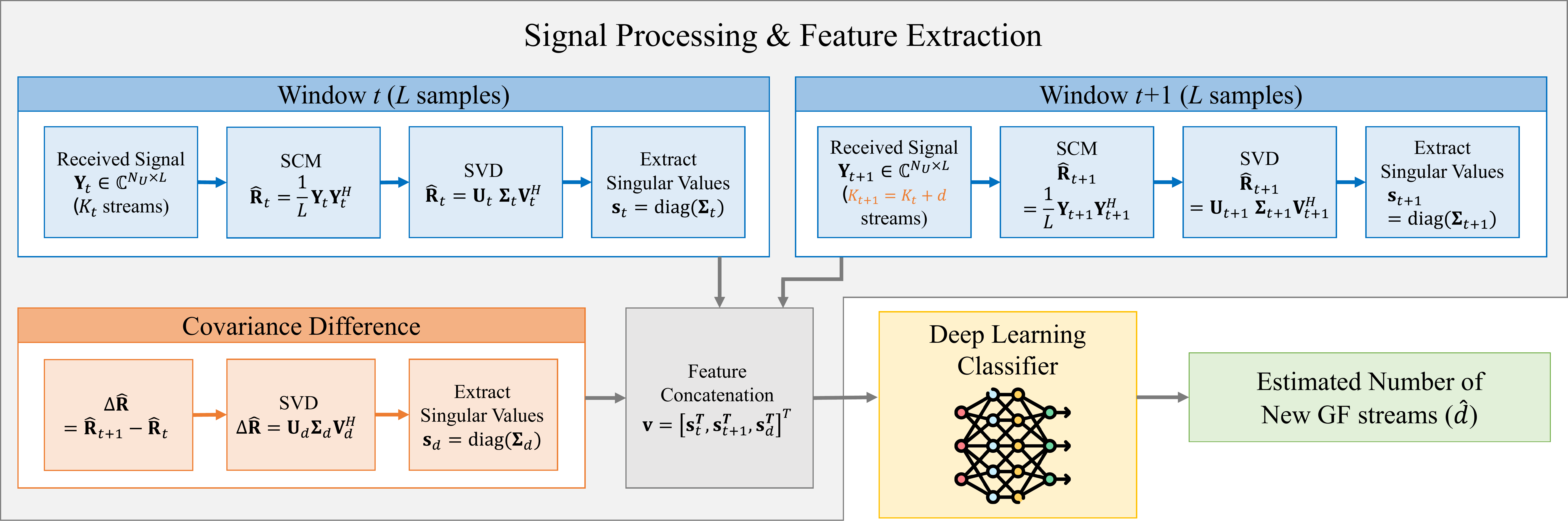}
    \caption{The proposed stream sensing method based on covariance differencing and a deep learning model.}
    \vspace{-3mm}
    \label{fig:proposed}
\end{figure*}
The main contributions of this paper are as follows:
\begin{itemize}
    \item \textbf{Covariance Differencing:} We propose a covariance differencing technique to isolate newly activated streams by subtracting the previous statistical state, suppressing interference from pre-existing transmissions.
    \item \textbf{Theoretical Analysis:} We rigorously analyze the sample covariance matrix (SCM) deviation induced by channel variations, establishing a theoretical bound based on channel correlation to determine the sensing window size.
    \item \textbf{DL Classifier:} We integrate a DL classifier to compensate for finite sampling errors, learning noise and stream statistics that exceed the limits of conventional methods.
\end{itemize}

\vspace{-2mm}
\section{System Model}
We consider an uplink MIMO system where a multi-antenna BS serves multiple UEs. As illustrated in Fig. \ref{fig:system_model}, the network accommodates the coexistence of two distinct user types: GB UEs and GF UEs. While GB transmissions are pre-scheduled by the BS, GF UEs operate sporadically, transmitting data without prior scheduling grants. In this system, we assume that up to $K_{GF}$ GF UEs are simultaneously active, sharing the uplink resources with ongoing GB transmissions.

To mitigate collisions, we employ an LBT strategy. Under this protocol, a potential transmitter first senses the channel to assess uplink occupancy and proceeds only if the channel is determined to be available. In our framework, a GF UE intending to transmit, referred to as the sensing UE, passively monitors the uplink environment to track current network activity. Equipped with $N_U$ antennas, the sensing UE is assumed to be positioned to reliably capture the aggregate energy of ongoing uplink transmissions. Leveraging this passive observation, the sensing UE estimates the instantaneous number of active streams to autonomously determine whether to proceed with transmission or defer access.

We specifically focus on the overloaded scenario where the number of active streams in sensing window $t$, denoted by $K_{t}$, is equal to or exceeds the number of receive antennas $N_U$ (i.e., $K_{t} \ge N_U$). In such scenario, conventional source enumeration methods fail to resolve the absolute number of streams. Therefore, our objective is to estimate the incremental change $d$ in the number of active streams between two consecutive observation windows.
To capture this variation, the sensing UE observes the uplink over two consecutive windows, $t$ and $t+1$, each consisting of $L$ samples. In window $t$, the received signal matrix $\mathbf{Y}_t \in \mathbb{C}^{N_U \times L}$ is modeled as:
\begin{equation}
    \mathbf{Y}_t = \mathbf{H}_t\mathbf{X}_t + \mathbf{N}_t,
\end{equation}
where $\mathbf{H}_t\in \mathbb{C}^{N_{U}\times K_t}$ represents the composite channel matrix from the $K_t$ active streams to the $N_{U}$ antennas of the sensing UE. $\mathbf{X}_t\in\mathbb{C}^{K_t\times L}$ is the matrix of transmitted signals, and $\mathbf{N}_t\in\mathbb{C}^{N_{U}\times L}$ denotes the additive white Gaussian noise (AWGN) matrix following $\mathcal{CN}(0, \sigma_t^2\mathbf{I})$.

Based on the observed signal, the sensing UE computes the SCM for window $t$ as:
\begin{equation}
    \hat{\mathbf{R}}_t = \frac{1}{L}\mathbf{Y}_t\mathbf{Y}_t^H,
    \label{eq:SCM}
\end{equation}
where $(\cdot)^H$ denotes the conjugate transpose.
Subsequently, in window $t+1$, we consider a scenario where $d$ new GF streams become active, resulting in a new total of $K_{t+1}=K_t + d$ streams. The sensing UE captures $\mathbf{Y}_{t+1}$ and calculates the corresponding SCM, $\hat{\mathbf{R}}_{t+1}$, following the same formulation. 
The proposed approach leverages the covariance difference between the two windows to isolate the information of the newly activated $d$ streams. By computing the difference matrix:
\begin{equation}
    \Delta \hat{\mathbf{R}}_t = \hat{\mathbf{R}}_{t+1} - \hat{\mathbf{R}}_t,
\end{equation}
We theoretically cancel out the spatial covariance components corresponding to the pre-existing streams. Consequently, the subspace of this difference matrix $\Delta \hat{\mathbf{R}}_t$ is primarily spanned by the new $d$ streams, effectively mitigating the influence of the $K_t$ existing signals. To analyze this new subspace and extract spectral features, a singular value decomposition (SVD) is performed:
\begin{equation}
    \Delta \hat{\mathbf{R}}_t = \mathbf{U}_d\mathbf{\Sigma}_d\mathbf{V}_d^H,
\end{equation}
where the singular values in the diagonal matrix $\mathbf{\Sigma}_d$ serve as features characterizing the newly activated streams. Similarly, SVD is applied to the individual SCMs to obtain $\hat{\mathbf{R}}_t = \mathbf{U}_t\mathbf{\Sigma}_t\mathbf{V}_t^H$ and $\hat{\mathbf{R}}_{t+1} = \mathbf{U}_{t+1}\mathbf{\Sigma}_{t+1}\mathbf{V}_{t+1}^H$.  

Ideally, the rank of $\Delta \hat{\mathbf{R}}_t$ would directly reveal $d$. However, in practice, finite sampling introduces estimation errors and residual noise, making conventional rank-based methods unreliable. To address this, we design a DL classifier, denoted as $f(\cdot)$, that estimates the newly activated streams $d$ using spectral features. We construct the model input $\mathbf{v} \in \mathbb{R}^{3N_U}$ by concatenating the singular value vectors from the respective SCMs: $\mathbf{v} = [\mathbf{s}_t^T, \mathbf{s}_{t+1}^T, \mathbf{s}_d^T]^T,$ where $\mathbf{s}_x = \text{diag}(\mathbf{\Sigma}_x)$ denotes the vector of singular values. Therefore, the estimation problem is formulated as $\hat{d} = f(\mathbf{v})$. Based on this result, the sensing UE updates its total stream count, providing a reliable basis for deciding whether to proceed with GF access.

\vspace{-2mm}
\section{Proposed Sensing Method}
As detailed in Fig.~\ref{fig:proposed} and Algorithm \ref{alg:proposed_sensing}, the method consists of SCM construction, covariance differencing for feature extraction, and DL-based classification.

\vspace{-3mm}
\subsection{SCM Deviation Analysis}
\label{subsec:scm_block}
\begin{figure}[h]
    \vspace{-3mm}
    \centering
    \includegraphics[width=0.8\linewidth]{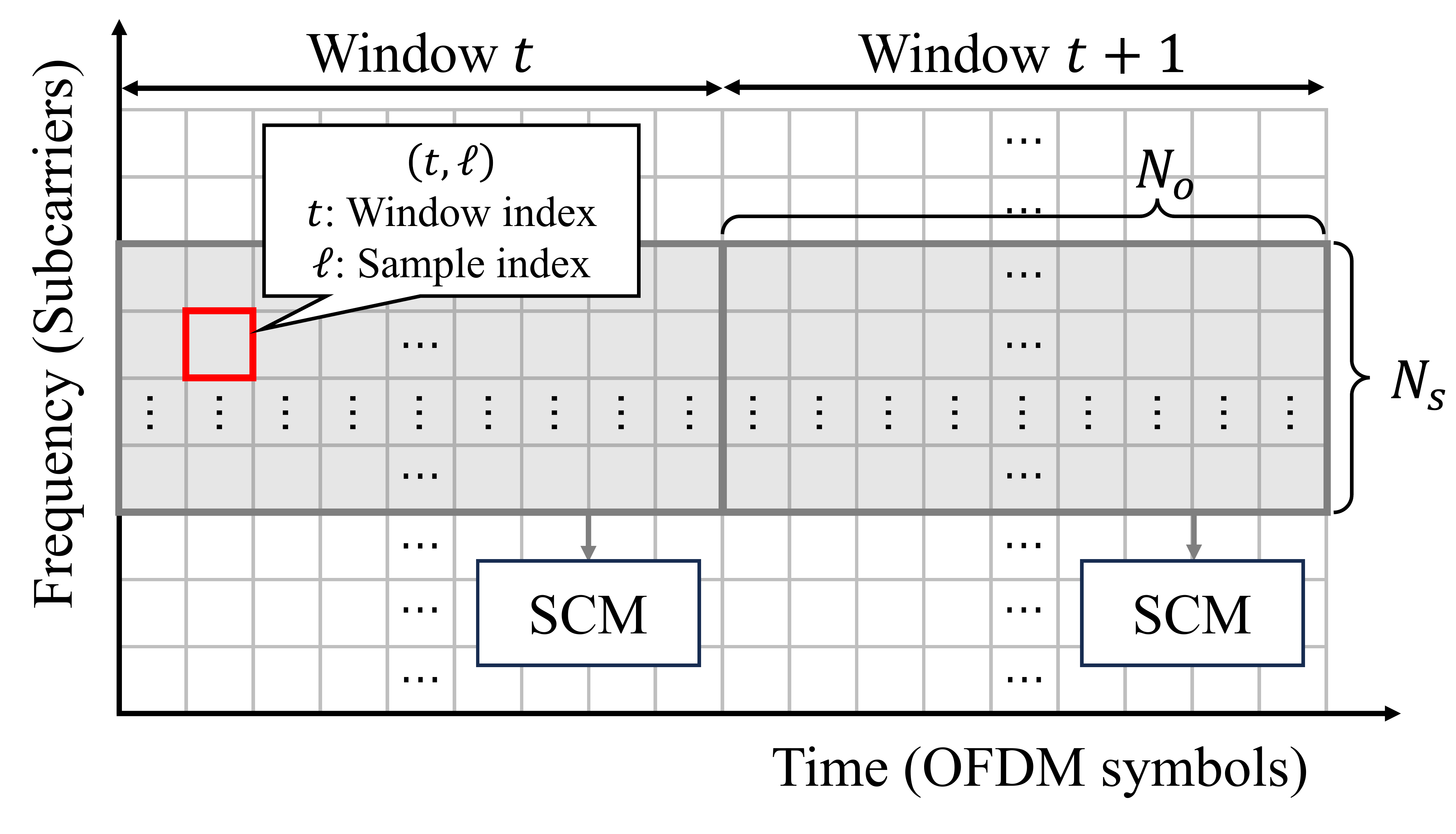}
    \caption{Time-frequency resource grid structure for sampling two consecutive sensing windows.}
    \vspace{-3mm}
    \label{fig:sampling_grid}
\end{figure}

The sensing UE constructs an SCM by aggregating samples over a time-frequency window. As depicted in Fig. \ref{fig:sampling_grid}, each window collects $L = N_o N_s$ samples, where $N_o$ and $N_s$ denote the number of OFDM symbols and subcarriers, respectively.

The aggregation of these distributed samples into an SCM is valid only if the channel remains sufficiently coherent across the window.
Therefore, we mathematically justify this approach by bounding the estimation error induced by channel variations.
The corresponding received signal vector for sample $\ell$ in window $t$ is modeled as:
\begin{equation}
    \mathbf y_{t,\ell}
    =
    \mathbf H_{t,\ell}\mathbf x_{t,\ell}
    + \mathbf n_{t,\ell},
    \quad
    \mathbf y_{t,\ell} \in\mathbb C^{N_U}, \quad \ell=1,\dots, L,
    \label{eq:ytl_model}
\end{equation}
where $\mathbf{H}_{t,\ell}\in\mathbb{C}^{N_U\times K_t}$ represents the channel matrix, and $\mathbf{x}_{t,\ell}$ and $\mathbf{n}_{t,\ell}$ denote the transmitted signal and AWGN vector, respectively.
Then, we stack the $L$ samples in window $t$ as $\mathbf{Y}_t = [\mathbf{y}_{t,1},\dots,\mathbf{y}_{t,L}] \in\mathbb{C}^{N_U\times L}$.

\noindent\textbf{Local stationarity:}
To quantify the impact of channel variation, we define a reference channel $\mathbf{H}_t$ (e.g., the channel response at the first sample $\ell=1$).
The instantaneous channel at sample $\ell$ is expressed as a perturbation from this reference:
\begin{equation}
    \mathbf H_{t,\ell}
    =
    \mathbf H_{t}
    + \Delta\mathbf H_{t,\ell},
    \label{eq:H_perturb}
\end{equation}
where $\Delta\mathbf H_{t,\ell}$ captures the residual time–frequency variation.

Let $\mathbf R_t = \frac{1}{L} \mathbb{E}[\mathbf{Y}_t\mathbf{Y}_t^H]$ denote the true statistical covariance matrix.
For simplicity, assume that $\mathbf x_{t,\ell}$ is zero-mean with covariance
$\mathbf R_x=\mathbb E[\mathbf x_{t,\ell}\mathbf x_{t,\ell}^H]$ that does not vary within the window.
Then, we obtain:
\begin{align}
    \mathbf R_t
    &= \frac{1}{L}\sum_{\ell=1}^{L}
       \Big(
         \mathbf H_{t,\ell}\mathbf R_x\mathbf H_{t,\ell}^H
         + \sigma_t^2\mathbf I_{N_U}
       \Big), \\
    \mathbf R_t^{\star}
    &= \mathbf H_t\mathbf R_x\mathbf H_t^H
       + \sigma_t^2\mathbf I_{N_U},
\end{align}
where $\mathbf R_t^{\star}$ is the reference covariance that would be observed if the channel is $\mathbf H_{t,\ell}=\mathbf H_t$ for all $\ell$.

The difference between the true statistical and the reference covariance is given by:
\begin{align}
\mathbf R_t - \mathbf R_t^{\star}
&= \frac{1}{L}\sum_{\ell=1}^{L} \Big(
      \mathbf H_t\mathbf R_x\Delta\mathbf H_{t,\ell}^H 
    + \Delta\mathbf H_{t,\ell}\mathbf R_x\mathbf H_t^H  \notag\\
&\qquad\qquad
    + \Delta\mathbf H_{t,\ell}\mathbf R_x\Delta\mathbf H_{t,\ell}^H
    \Big).
\label{eq:Rt_diff_expansion}
\end{align}
By applying the triangle inequality and the submultiplicativity of matrix norms, we obtain:
\begin{equation}
\begin{split}
\big\|\mathbf R_t - \mathbf R_t^{\star}\big\|_F \\
\le \frac{\|\mathbf R_x\|_2}{L}&
   \sum_{\ell=1}^{L}
   \left(
      2\|\mathbf H_t\|_F\,
        \|\Delta\mathbf H_{t,\ell}\|_F
      + \|\Delta\mathbf H_{t,\ell}\|_F^2
   \right).
\label{eq:cov_mismatch_general_bound}
\end{split}
\end{equation}
where $\|\cdot\|_2$ denotes the spectral norm.

\noindent\textbf{Channel correlation and covariance bound:}
We derive a correlation-dependent upper bound on this deviation.
Let $\rho_h(\ell,\ell')$ denote the normalized correlation coefficient between samples $\ell$ and $\ell'$, defined as:
\begin{equation}
    \rho_h(\ell,\ell')
    =
    \frac{\mathbb{E}\big[h_{t,\ell}\,h^{\ast}_{t,\ell'}\big]}{\sqrt{\mathbb{E}\big[|h_{t,\ell}|^2\big]\,\mathbb{E}\big[|h_{t,\ell'}|^2\big]}},
    \label{eq:rho_s_def}
\end{equation}
where $h_{t,\ell} = [H_{t,\ell}]_{m,k}$ denotes the $(m,k)$-th element of the channel matrix at sample index $\ell$.
We choose the window size such that the channel samples within a window remain sufficiently coherent, i.e., their pairwise correlations are positive:
\begin{equation}
    \Re\{\rho_h(\ell,\ell')\} \ge \rho_{\mathrm{th}},
    \quad
    \forall\, \ell,\ell'\in\{1,\dots,L\},
    \label{eq:rho_s_thresh}
\end{equation}
for a prescribed correlation threshold $\rho_{\mathrm{th}}\in(0,1)$, where $\Re\{\cdot\}$ denotes the real part.
For two samples $\ell$ and $\ell'$ satisfying~\eqref{eq:rho_s_thresh}, and under the assumption $\mathbb{E}\!\left[|h_{t,\ell}|^2\right]=\mathbb{E}\!\left[|h_{t,\ell'}|^2\right]$, we obtain:

\begin{align}
    \mathbb{E}\!\left[|h_{t,\ell'} - h_{t,\ell}|^2\right]
    &= 2\,\mathbb{E}[|h_{t,\ell}|^2]\big(1-\Re\{\rho_h(\ell,\ell')\}\big) \nonumber\\
    &\le 2\,\mathbb{E}[|h_{t,\ell}|^2]\big(1-\rho_{\mathrm{th}}\big).
\end{align}
Aggregating this bound over all entries of the channel matrix and using the Frobenius norm, we obtain:
\begin{equation}
    \mathbb{E}\big[\|\Delta\mathbf H_{t,\ell}\|_F^2\big]
    \;\le\;
    2\big(1-\rho_{\mathrm{th}}\big)\,
    \mathbb{E}\big[\|\mathbf H_{t}\|_F^2\big],
    \label{eq:deltaH_rho_relation}
\end{equation}
for any sample within a correlation-coherent window.
Consequently, the mean-square perturbation can be bounded as:
\begin{equation}
    \mathbb{E}\big[\|\Delta\mathbf H_{t,\ell}\|_F^2\big]
    \le \varepsilon_H^2,
    \quad
    \varepsilon_H =
    \sqrt{2\big(1-\rho_{\mathrm{th}}\big)}\,
    \sqrt{\mathbb{E}\big[\|\mathbf H_t\|_F^2\big]}.
    \label{eq:epsH_def}
\end{equation}
Finally, taking the expectation of \eqref{eq:cov_mismatch_general_bound} and applying the Cauchy-Schwarz inequality, we obtain the theoretical bound:
\begin{equation}
    \begin{split}
    \mathbb{E} \big[\big\|\mathbf R_t - \mathbf R_t^{\star}\big\|_F \big]
    &\le
    \|\mathbf R_x\|_2
    \Big(
       2\sqrt{\mathbb{E}[\|\mathbf{H}_t\|_F^2]}\varepsilon_H
       + \varepsilon_H^2
    \Big) \\
    &=
    \mathcal{O}\!\left(\sqrt{1-\rho_{\mathrm{th}}}\right),
    \label{eq:cov_mismatch_rho_bound}
    \end{split}
\end{equation}
where the dominant term scales with $\sqrt{1-\rho_{\mathrm{th}}}$.
This bound implies that as long as the correlation threshold $\rho_{\mathrm{th}}$ is chosen sufficiently high, the channel remains effectively coherent within a window and $\mathbf{R}_t$ stays close to the reference covariance $\mathbf{R}_t^{\star}$. Therefore, with a high $\rho_{\mathrm{th}}$, we can construct the SCM from $L$ distributed samples within the window while keeping the deviation due to channel variation negligible.

\vspace{-3mm}
\subsection{Covariance Differencing}
The goal is to extract informative features by isolating the signal subspace of the newly activated $d$ streams.
The theoretical basis for this approach is derived using the reference static covariance, $\mathbf{R}_t^{\star}$ established in Section~\ref{subsec:scm_block}.
To facilitate the derivation, we adopt the following assumptions:
\begin{itemize}
    \item The transmitted signal $\mathbf{x}_t$ and additive noise $\mathbf{n}_t$ are zero-mean and statistically independent.
    \item  The transmitted symbols are independent and identically distributed (i.i.d.) with $\mathbb{E}[\mathbf{x}_t\mathbf{x}_t^{H}]=\mathbf{I}$.
    \item The channels for the $K_t$ pre-existing streams are assumed to be static across the two observation windows.
\end{itemize}
Then, the reference covariance $\mathbf{R}_t^{\star}$ is expressed as:
\begin{equation}
\begin{aligned}
   \mathbf{R}_t^{\star} = \mathbf{H}_t\mathbf{H}_t^H + \sigma_t^2\mathbf{I}_{N_U}
    = \sum_{k=1}^{K_t} \mathbf{h}_k\mathbf{h}_k^H + \sigma_t^2\mathbf{I}_{N_U},
\end{aligned}
\label{eq:Rt_star}
\end{equation}
where $\mathbf{H}_t = [\mathbf{h}_1, \dots, \mathbf{h}_{K_t}]$ is the channel matrix with $\mathbf{h}_k$ representing the channel vector for the $k$-th stream.
Following the same logic, the reference covariance for the next window $t+1$, with $d$ newly added streams, is:
\begin{equation}
    \mathbf{R}_{t+1}^{\star} = \sum_{k=1}^{K_{t}+d} \mathbf{h}_k\mathbf{h}_k^H + \sigma_{t+1}^2\mathbf{I}_{N_U},
    \label{eq:Rt_star_next}
\end{equation}
where the channel matrix $\mathbf{H}_{t+1}$, now includes the newly added $d$ streams, $\mathbf{H}_{t+1} = [\mathbf{h}_1, \dots,\mathbf{h}_{K_t},\mathbf{h}_{K_{t}+1},\dots, \mathbf{h}_{K_{t}+d}]$.
Subtracting \eqref{eq:Rt_star} from \eqref{eq:Rt_star_next} yields the ideal difference matrix:
\begin{equation}
    \Delta\mathbf{R}_t^{\star} = \mathbf{R}_{t+1}^{\star} - \mathbf{R}_{t}^{\star} = \sum_{k=K_t + 1}^{K_{t}+d} \mathbf{h}_k\mathbf{h}_k^H+ (\sigma_{t+1}^2 - \sigma_{t}^2) \mathbf{I}_{N_U}.
    \label{eq:delta_star}
\end{equation}
Under the assumption of stationary noise power across consecutive windows, this theoretical result demonstrates that the dominant subspace of $\Delta \mathbf{R}_t^{\star}$ is determined solely by the $d$ newly activated streams, theoretically free from the interference of pre-existing users.

In practice, we apply SVD to $\hat{\mathbf{R}}_t$, $\hat{\mathbf{R}}_{t+1}$, and $\Delta\hat{\mathbf{R}}_t$ to extract spectral features.
The resulting singular value vectors, $\mathbf{s}_t$, $\mathbf{s}_{t+1}$, and $\mathbf{s}_d$, characterize the spectral profiles of the pre-existing, total, and newly activated streams, respectively.
These vectors are concatenated into a vector $\mathbf{v}$ to feed the DL classifier.

\begin{algorithm}[t]
\caption{Proposed Stream Sensing Method}
\label{alg:proposed_sensing}
\begin{algorithmic}[1]
\State \textbf{Input:} $\mathbf{Y}_t$, $\mathbf{Y}_{t+1}$, $L$, $f(\cdot)$
\For{$i \in \{t, t+1\}$} \Comment{Step 1: SCM Construction}
    \State $\hat{\mathbf{R}}_{i} \gets \frac{1}{L}\mathbf{Y}_{i}\mathbf{Y}_{i}^{H}$
    \State $[\mathbf{U}_{i},\mathbf{\Sigma}_{i},\mathbf{V}_{i}^{H}] \gets \mathrm{SVD}(\hat{\mathbf{R}}_{i})$ 
    \State $\mathbf{s}_{i} \gets \mathrm{diag}(\mathbf{\Sigma}_{i})$
\EndFor
\State $\Delta\hat{\mathbf{R}}_t \gets \hat{\mathbf{R}}_{t+1} - \hat{\mathbf{R}}_{t}$ \Comment{Step 2: Covariance Differencing}
\State $[\mathbf{U}_{d},\mathbf{\Sigma}_{d},\mathbf{V}_{d}^{H}] \gets \mathrm{SVD}(\Delta\hat{\mathbf{R}}_t)$
\State $\mathbf{s}_{d} \gets \mathrm{diag}(\mathbf{\Sigma}_{d})$
\State $\mathbf{v} \gets [\mathbf{s}_{t}^{T}, \mathbf{s}_{t+1}^{T}, \mathbf{s}_{d}^{T}]^{T}$ \Comment{Step 3: DL Classification}
\State $\hat{d} \gets f(\mathbf{v})$
\State \textbf{Output:} $\hat{d}$
\end{algorithmic}
\end{algorithm}

\vspace{-3mm}
\subsection{Proposed DL Classifier}
The proposed classifier estimates the number of newly activated streams $d$ using a two-stream architecture as summarized in Table \ref{table:classifier_spec}. The input $\mathbf{v}$ is processed in parallel branches. 
Each branch comprises a sequence of a fully connected (FC) layer, batch normalization (BN), and rectified linear unit (ReLU) activation, and their outputs are fused for final classification. The network is trained using cross-entropy loss with the Adam optimizer. Inference yields the estimate $\hat{d}$ via:
\begin{equation}
\hat{d}=\arg\max_{k\in{0,\ldots,K_{\text{GF}}}} f(\mathbf{v})_k,
\end{equation}
where $f(\mathbf{v})_k$ denotes the $k$-th output logit.

\vspace{-2mm}
\begin{table}[h]
\centering
\caption{Architecture of the proposed DL classifier.}
\label{table:classifier_spec}
\begin{tabular}{@{}lll@{}}
\toprule
\textbf{Stage} & \textbf{Module / Input} & \textbf{Configuration} \\ \midrule
\multirow{2}{*}{\begin{tabular}[c]{@{}l@{}}Feature\\ Extraction\end{tabular}} & Stream 1 / $[\mathbf{s}_t^T, \mathbf{s}_{t+1}^T]^T$ & FC(32) $\rightarrow$ BN $\rightarrow$ ReLU \\ \cmidrule(l){2-3} 
 & Stream 2 / $\mathbf{s}_d$ & FC(32) $\rightarrow$ BN $\rightarrow$ ReLU \\ \midrule
Feature Fusion & Concatenation & Concatenate Stream Latents \\ \midrule
\multirow{2}{*}{Classification} & Fusion Head & FC(16) $\rightarrow$ BN $\rightarrow$ ReLU \\ \cmidrule(l){2-3} 
 & Output Layer & FC($K_{\text{GF}} + 1$) $\rightarrow$ Logits \\ \bottomrule
\end{tabular}
\vspace{-5mm}
\end{table}

\section{Simulation Results}
\subsection{Simulation Settings} \label{subsec:sim_setting}

\begin{figure}[h]
    \vspace{-2mm}
    \centering
    \includegraphics[width=0.8\linewidth]{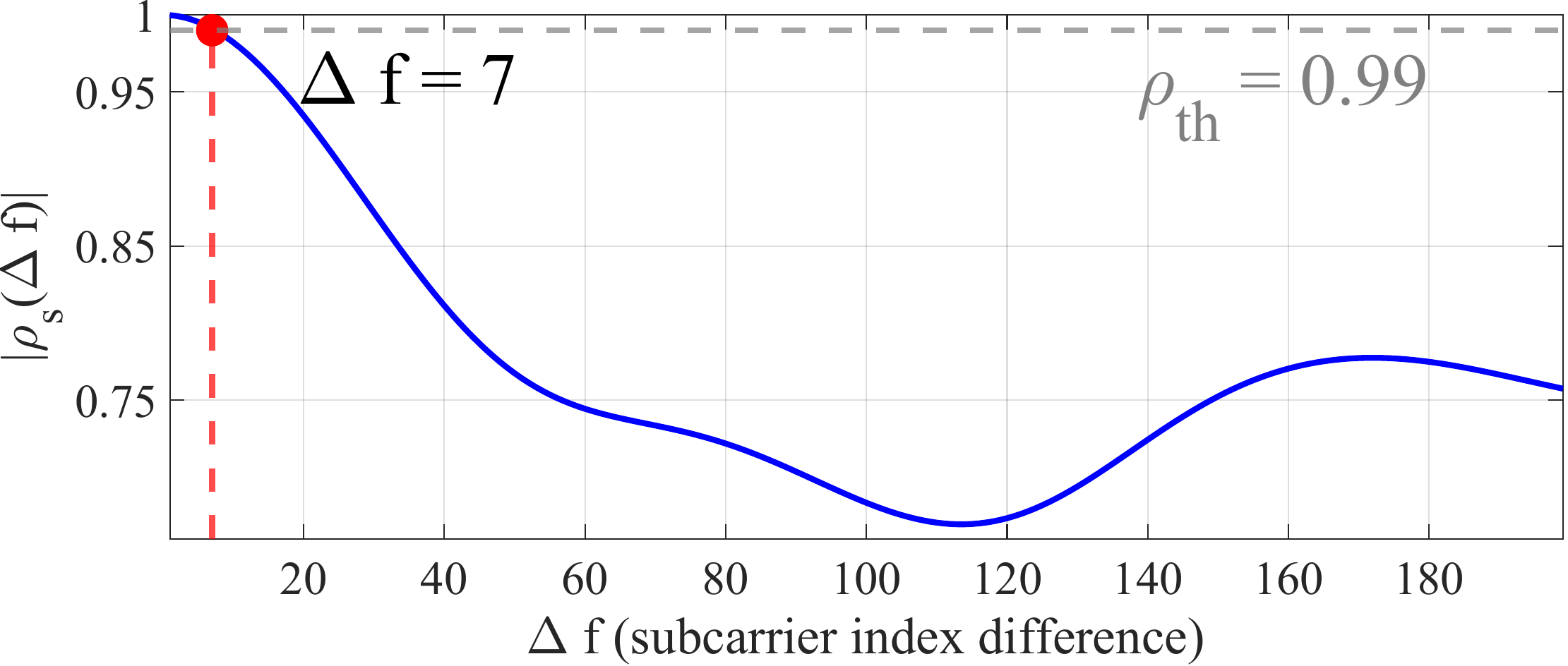}
    \caption{Frequency-domain correlation of the 3GPP TR 38.901 TDL-A channel versus subcarrier index difference $\Delta f$.}
    \vspace{-3mm}
    \label{fig:freq_corr}
\end{figure}

We assess performance under two channel models: (i) an i.i.d. flat Rayleigh channel generated as complex Gaussian fading, and (ii) a 3GPP TR 38.901 TDL-A channel model \cite{3GPP.38.901} implemented in \texttt{Sionna} \cite{hoydis2022sionna} with a 3.5~GHz carrier, 30~kHz subcarrier spacing, 2048-point FFT, and 100~ns delay spread.
Specifically for the 3GPP TR 38.901 TDL-A channel, we characterize the correlation defined in (\ref{eq:rho_s_def}), as a function of the subcarrier index difference $\Delta f$: 
\begin{equation}
\label{eq:rho_subcarrier_compact2}
\rho_h(\Delta f) =\mathbb{E}_l\!\big[\rho_h(l,\,l+\Delta f)\big],
\end{equation}
where $l$ denotes the subcarrier index. We set $N_s=7$ to ensure that the frequency-domain channel correlation $|\rho_h(\Delta f)|$ remains above $0.99$, as shown in Fig.~\ref{fig:freq_corr}. We also fix $N_o=140$ and assume that both the channel and noise power remain static over a frame, since its duration is short enough relative to the channel coherence time. Additionally, we consider $N_U=4$ receive antennas, where the number of newly activated streams $d$ is randomly selected up to a maximum of $K_{GF}=3$.

\vspace{-4mm}
\subsection{Baselines}
To evaluate the effectiveness of the proposed framework, we compare the performance of the following schemes:
\begin{itemize}
    \item \textbf{Proposed:} The complete framework utilizing the full concatenated vector $\mathbf{v} = [\mathbf{s}_t^T, \mathbf{s}_{t+1}^T, \mathbf{s}_d^T]^T$ to exploit both raw and differenced spectral information.
    \item \textbf{Difference Only:} An ablation baseline trained exclusively on the difference features $\mathbf{s}_d$.
    \item \textbf{Raw Only:} A baseline using only the raw features $[\mathbf{s}_t^T, \mathbf{s}_{t+1}^T]^T$ to assess detection performance in the absence of covariance differencing.
    \item \textbf{Thresholding:} A non-DL method that estimates $d$ by counting singular values in $\mathbf{s}_d$ exceeding a threshold which is empirically optimized on a validation set.
    \item \textbf{MDL:} A standard approach applying the MDL criterion to the singular values $\mathbf{s}_d$ of the difference matrix $\Delta\hat{\mathbf{R}}_t$.
\end{itemize}

\begin{figure}[t!]
    \centering
    \begin{subfigure}[t]{0.48\linewidth}
        \centering
        \includegraphics[width=\linewidth]{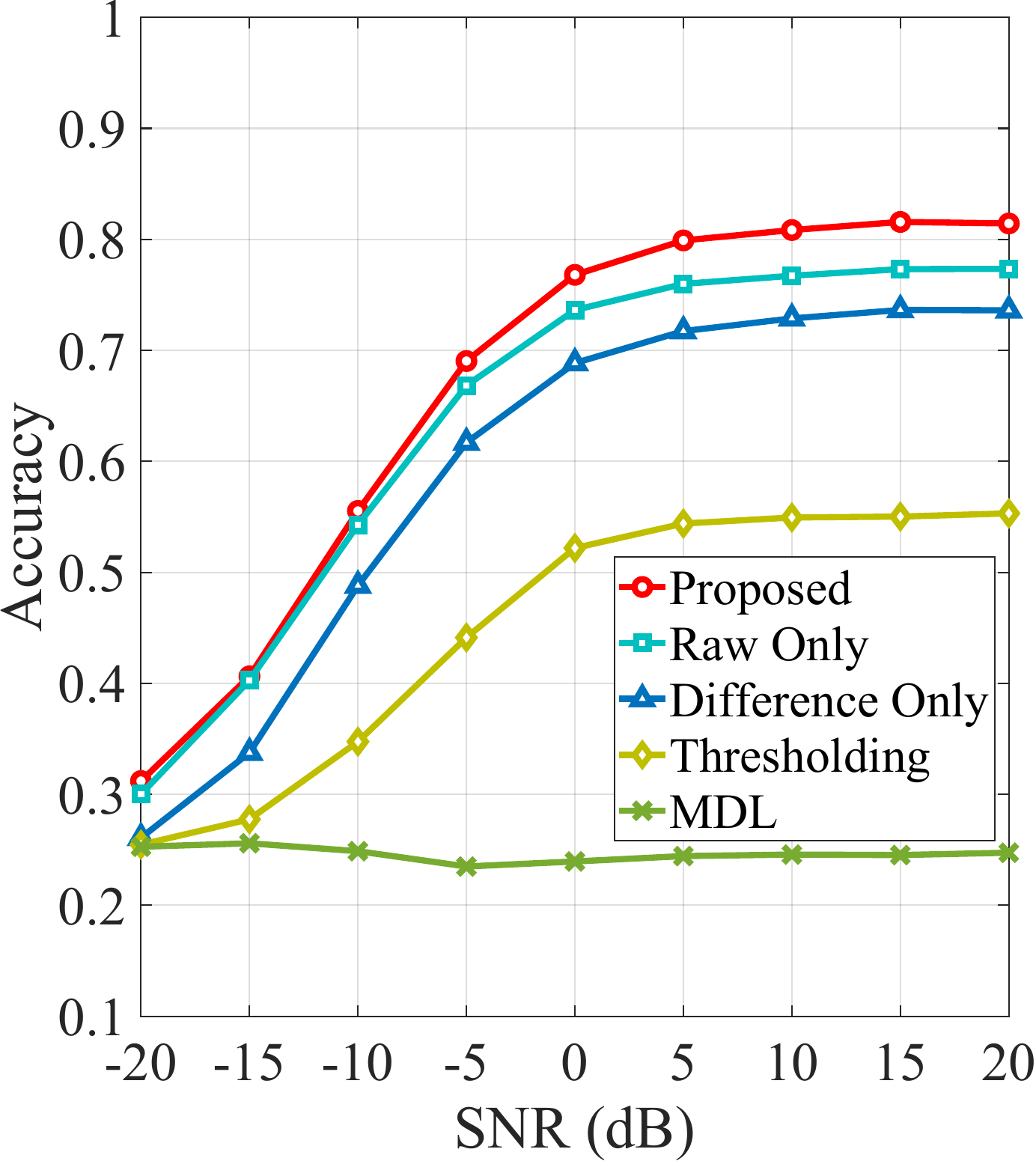}
        \caption{I.i.d. flat Rayleigh fading}
        \label{fig:flat_snr}
    \end{subfigure}
    \hfill
    \begin{subfigure}[t]{0.48\linewidth}
        \centering
        \includegraphics[width=\linewidth]{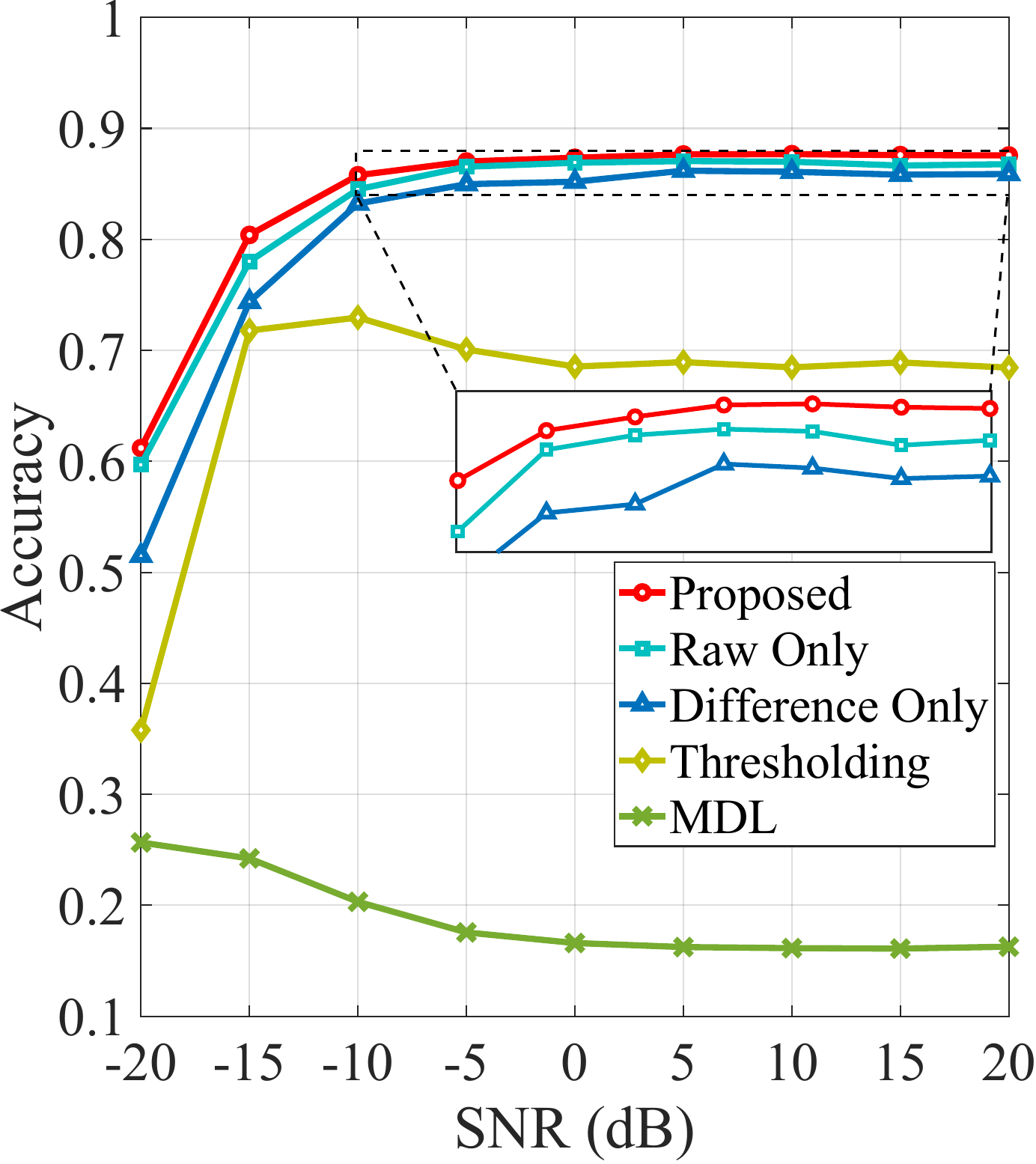}
        \caption{3GPP TR 38.901 TDL-A}
        \label{fig:3gpp_snr}
    \end{subfigure}
    \caption{Sensing accuracy versus the SNR with $K_t = 4$.}
    \label{fig:perf_compare_snr}
    \vspace{-3mm}
\end{figure}

\begin{figure}[t!]
    \centering
    \begin{subfigure}[t]{0.48\linewidth}
        \centering
        \includegraphics[width=\linewidth]{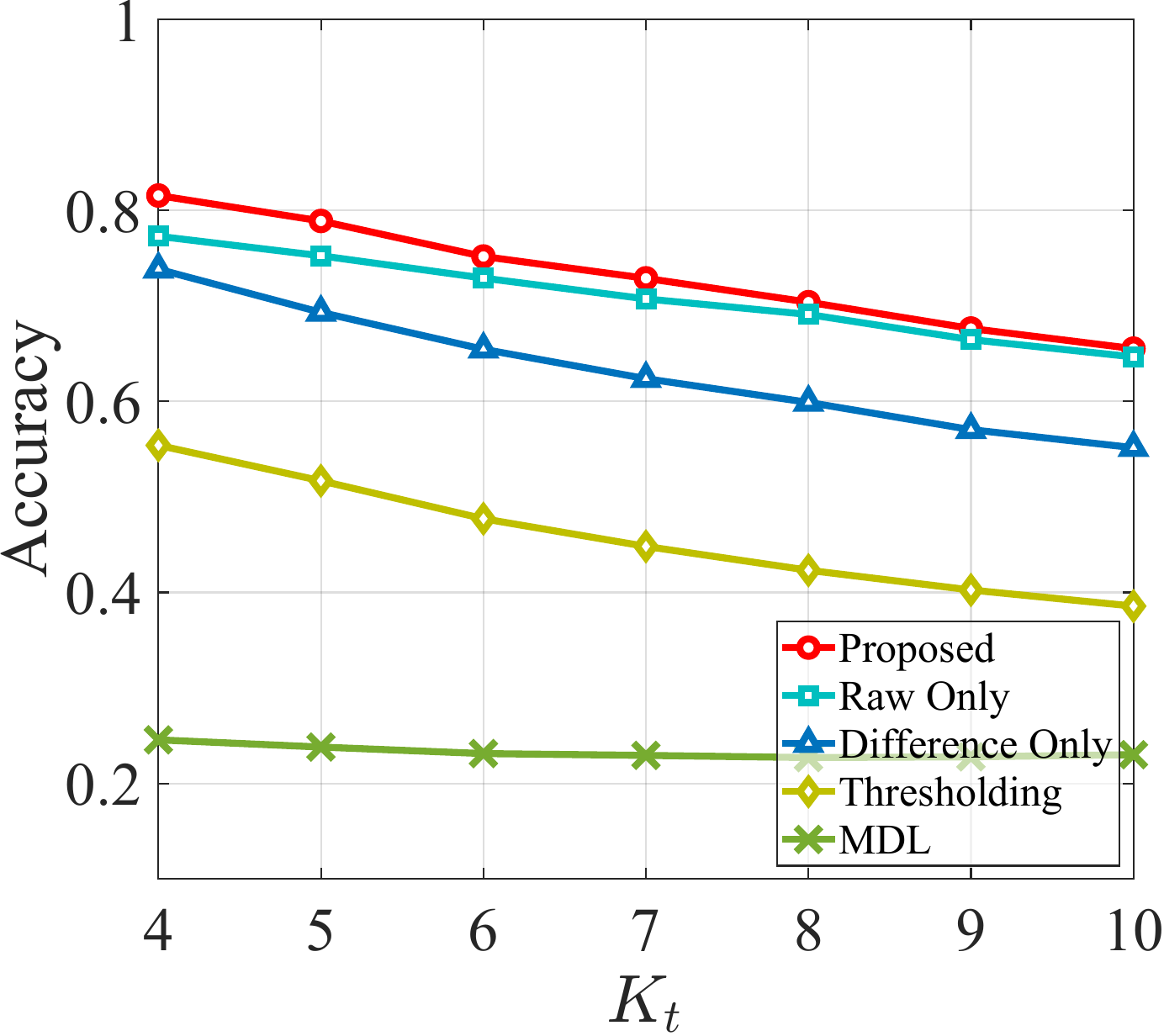}
        \caption{I.i.d. flat Rayleigh fading}
        \label{fig:flat_gb}
    \end{subfigure}
    \hfill
    \begin{subfigure}[t]{0.48\linewidth}
        \centering
        \includegraphics[width=\linewidth]{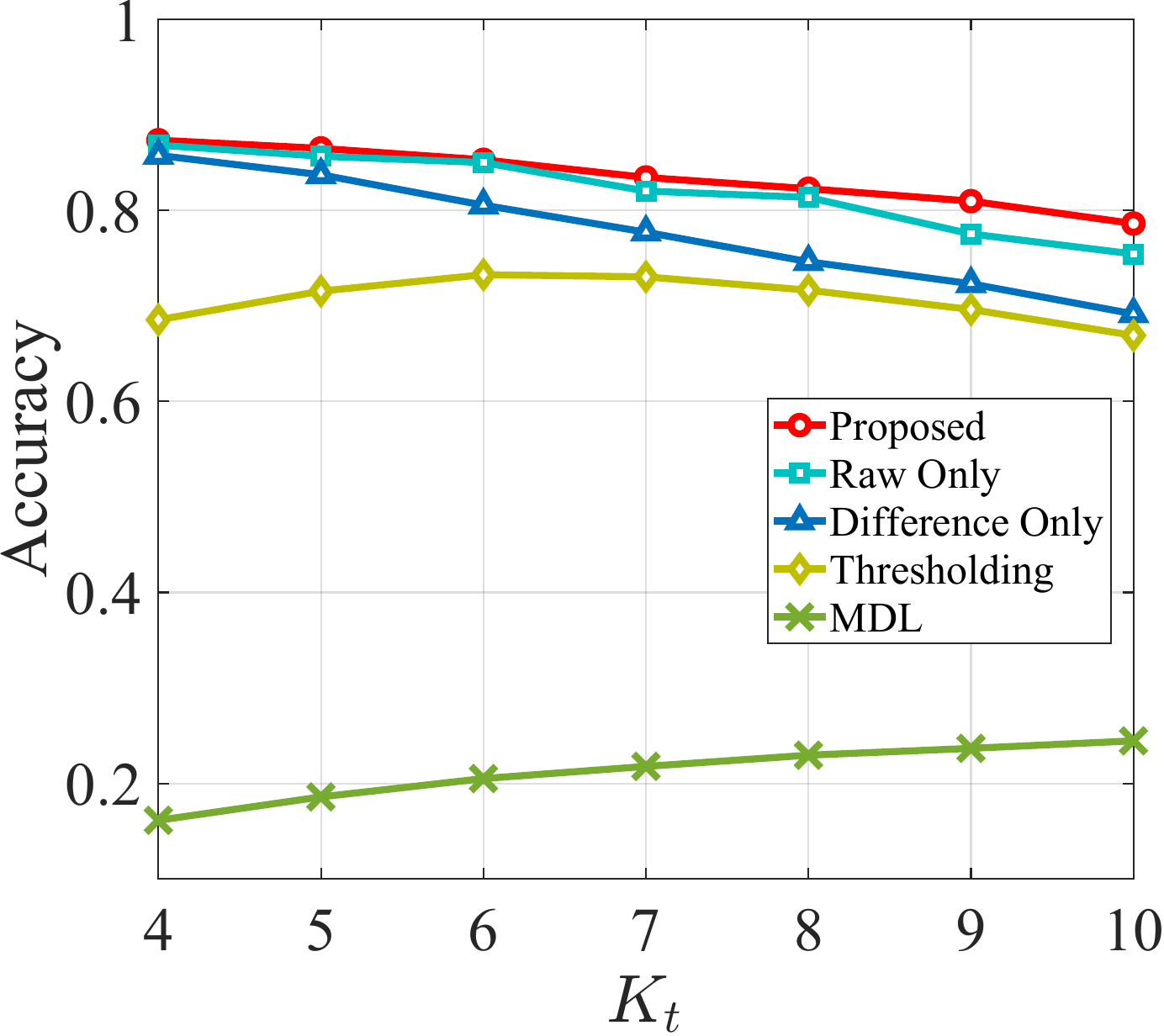}
        \caption{3GPP TR 38.901 TDL-A}
        \label{fig:3gpp_gb}
    \end{subfigure}
    \caption{Sensing accuracy versus the number of pre-existing streams $K_t$ at SNR = 20 dB.}
    \label{fig:perf_compare_gb}
    \vspace{-3mm}
\end{figure}

\vspace{-4mm}
\subsection{Results}
In all simulations, the sensing accuracy is defined as the probability that the estimated number of new streams $\hat{d}$ exactly matches the true value $d$. 
As illustrated in Fig.~\ref{fig:perf_compare_snr}, the \textbf{Proposed} method consistently achieves the highest accuracy across the entire tested SNR range for both channel models.
In the i.i.d. flat Rayleigh fading scenario, the \textbf{Proposed} scheme significantly outperforms the non-DL baselines. Specifically, \textbf{MDL} exhibits accuracy close to random guessing ($\sim 0.25$) due to the indistinguishable spectral gap in the overloaded regime.
Notably, the \textbf{Proposed} scheme surpasses both the \textbf{Raw Only} and \textbf{Difference Only} baselines. This confirms that relying solely on differencing or raw features is insufficient. The fusion of these features allows the DL classifier to effectively learn the signal subspace structure, yielding a performance gain of approximately 5.5\% and 10.8\% over the \textbf{Raw Only} and \textbf{Difference Only} baselines, respectively, at SNR = 20 dB.
Furthermore, in the 3GPP TR 38.901 TDL-A channel environment, the \textbf{Proposed} method maintains its superiority with a peak accuracy of 0.877. This robustness is attributed to our aggregation strategy, which groups 7 consecutive subcarriers, as verified in Section \ref{subsec:sim_setting}. This localized grouping ensures high intra-window correlation by keeping the frequency span within the coherence bandwidth. 
Consequently, this approach effectively increases the sample size for SCM estimation without suffering from channel variations, enabling reliable detection even under frequency-selective fading conditions.

Fig.~\ref{fig:perf_compare_gb} investigates the impact of the number of pre-existing streams $K_t$. As $K_t$ increases, the interference from existing users intensifies, causing the system to become increasingly overloaded. Despite this challenging environment, the \textbf{Proposed} scheme consistently demonstrates superior accuracy compared to the baselines across all tested $K_t$ values. This indicates that covariance differencing mitigates interference, and feature fusion ensures detection robustness against noise.

\section{Conclusion}
We proposed a robust stream sensing framework for overloaded GF access by leveraging covariance differencing to isolate newly activated streams. To address residual estimation errors inherent in finite-sample regimes, we integrated a DL classifier that fuses raw and differential spectral features to exploit their complementary information. Simulation results demonstrate that this approach significantly outperforms conventional methods and maintains high detection accuracy even in realistic frequency-selective fading conditions.

\bibliographystyle{IEEEtran}
\bibliography{refs}

\end{document}